\documentclass[twocolumn]{revtex4}
\usepackage[dvips]{graphicx}
\usepackage{dcolumn}
\usepackage{bm}
\usepackage{latexsym}
\begin{document}
\title{
Extended floor field CA model for evacuation dynamics
}
\author{Katsuhiro Nishinari}
\email{knishi@rins.ryukoku.ac.jp}
\affiliation{%
Department of Applied Mathematics and Informatics, 
Ryukoku University, Shiga, Japan 
}%
\author{Ansgar Kirchner}%
 \email{aki@thp.uni-koeln.de}
\affiliation{%
Institut  f\"ur Theoretische  Physik, Universit\"at 
zu K\"oln D-50937 K\"oln, Germany
}%
\author{Alireza Namazi}%
 \email{an@thp.uni-koeln.de}
\affiliation{%
Institut  f\"ur Theoretische  Physik, Universit\"at 
zu K\"oln D-50937 K\"oln, Germany
}%
\author{Andreas Schadschneider}%
 \email{as@thp.uni-koeln.de}
\affiliation{%
Institut  f\"ur Theoretische  Physik, Universit\"at 
zu K\"oln D-50937 K\"oln, Germany
}%
\date{\today}
\begin{abstract}
The floor field model, which is a cellular automaton model for
studying evacuation dynamics, is investigated and extended. A
method for calculating the static floor field, which describes the
shortest distance to an exit door, in an arbitrary geometry of
rooms is presented. The wall potential and contraction effect at a
wide exit are also proposed in order to obtain realistic behavior
near corners and bottlenecks. These extensions are important for
evacuation simulations, especially in the case of panics.
\end{abstract}
\maketitle
\section{Introduction}\label{intro}
Recent progress in modelling pedestrian dynamics \cite{book} is
remarkable and many valuable results are obtained by using
different models, such as the social force model \cite{HFV} and
the floor field model \cite{BKAJ,AA}. The former model is based on
a system of coupled differential equations which has to be solved
e.g.\ by using a molecular dynamics approach similar to the study
of granular matter. Pedestrian interactions are modelled via
long-ranged repulsive forces. In the latter model two kinds of
floor fields, i.e., a static and a dynamic one, are introduced to
translate a long-ranged spatial interaction into an attractive
local interaction, but with memory, similar to the phenomenon of
chemotaxis in biology \cite{chemo}. It is interesting that, even
though these two models employ different rules for pedestrian
dynamics, they share many properties including lane formation,
oscillations of the direction at bottlenecks \cite{BKAJ}, and the
so-called faster-is-slower effect \cite{HFV}. Although these are
important basics for pedestrian modelling, there are still many
things to be done in order to apply the models to more practical
situations such as evacuation from a building with complex
geometry.

In this paper, we will propose a method to construct the static
floor field for complex rooms of {\it arbitrary} geometry. The
static floor field is an important ingredient of the model and has
to be specified before the simulations. Moreover, the effect of
walls and contraction at a wide exit will be taken into account
which enables us to obtain realistic behavior in evacuation
simulations even for the case of panic situations.

This paper is organized as follows. In Sec.~\ref{humandata}, we
cite experimental data of evacuations to illustrate the strategy
of people in panic situations. Then an extended floor field model
is introduced in Sec.~\ref{CAmodel} including a method of
constructing the floor field and wall potentials. In
Sec.~\ref{simulation} results of simulations for various
configurations of a room are investigated and concluding
discussions are given in Sec.~\ref{conc}.

\section{Human behaviour in panic situations}\label{humandata}
First we discuss the different kinds of human behavior in panic
situations. People in a room try to evacuate in case of fire with
their own strategy. The strategies of the evacuation are well
studied up to now, and we cite an example of an experiment of
evacuation that was conducted in a large supermarket in Japan
\cite{Abe}. Fire alarms and false smoke were set suddenly in the
experiment, and after people had escaped from the building they
have been interviewed about their choice of escape routes etc.
Data from more than 300 people were collected. The following list
shows the statistics of the answers given:
\begin{enumerate}
 \item
    I escaped according to the signs and instructions, and
    also broadcast or guide by shopgirls (46.7\%).
 \item
    I chose the opposite direction to the smoking area
    to escape from the fire as soon as possible (26.3\%).
 \item
    I used the door because it was the nearest one (16.7\%).
 \item
    I just followed the other persons (3.0\%).
 \item
    I avoided the direction where many other persons go (3.0\%).
 \item
    There was a big window near the door and you could see outside.
    It was the most ``bright'' door, so I used it (2.3\%).
 \item
    I chose the door which I'm used to (1.7\%).
\end{enumerate}
We see that very different, sometimes even contradictory, choices
were made indicating the complexity of an evacuation problem. If
we assume that there are no signs and no guidance by broadcasts as
well as no information about the location of the fire, then
according to the questionnaires, people will try to evacuate by
relying on both one's memory of the route to the nearest door and
other people's behavior. This competition between collective and
individual behavior is essential for modelling evacuation
phenomena. It is included in the {\em static and dynamic floor
fields} of our model that we have introduced in previous papers
\cite{BKAJ,AA,AKA}.

\section{An extended floor field model}\label{CAmodel}
\begin{figure}[tb]
\begin{center}
\includegraphics[width=0.35\textwidth]{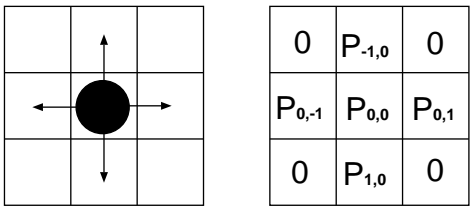}
\end{center}
\caption{ Target cells for a person at the next time step. The von
Neumann neighborhood is used for this model. } \label{figmove}
\end{figure}

In this section we will summarize the update rules of an extended
floor field model for modelling panic behavior of people
evacuating from a room. The space is discretized into cells of
size $40\,{\rm cm} \times 40\,{\rm cm}$ which can either be empty
or occupied by one pedestrian ({\em hard-core-exclusion}). Each
pedestrian can move to one of the unoccupied next-neighbor cells
$(i,j)$ (or stay at the present cell) at each discrete time step
$t\to t+1$ according to certain transition probabilities $p_{ij}$
(Fig.~\ref{figmove}) as explained below in Sec.~\ref{subrules}.

For the case of evacuation processes, the {\em static floor field}
$S$ describes the shortest distance to an exit door. The field
strength $S_{ij}$ is set inversely proportional to the distance
from the door. The {\em dynamic floor field} $D$ is a {\em virtual
trace} left by the pedestrians similar to the pheromone in
chemotaxis \cite{chemo}. It has its own dynamics, namely diffusion
and decay, which leads to broadening, dilution and finally
vanishing of the trace. At $t=0$ for all sites $(i,j)$ of the
lattice the dynamic field is zero, i.e., $D_{ij}=0$. Whenever a
particle jumps from site $(i,j)$ to one of the neighboring cells,
$D$ at the origin cell is increased by one.

The model is able to reproduce various fundamental phenomena, such
as lane formation in a corridor, herding and oscillation at a
bottleneck \cite{BKAJ,AA}. This is an indispensable property for
any reliable model of pedestrian dynamics, especially for
discussing safety issues.
\subsection{Basic update rules}\label{subrules}
The update rules of our CA have the following structure:
\begin{enumerate}
\item The dynamic floor field $D$ is modified according to its
diffusion and decay rules, controlled by the parameters $\alpha$
and $\delta$. In each time step of the simulation each single
boson of the whole dynamic field $D$ decays with probability
$\delta$ and diffuses with probability  $\alpha$ to one of its
neighboring cells.

\item For each pedestrian, the transition probabilities $p_{ij}$
for a move to an unoccupied neighbor cell $(i,j)$ are determined
by the two floor fields and one's inertia (Fig.~\ref{figmove}).
The values of the fields $D$ (dynamic) and $S$ (static) are
weighted with two sensitivity parameters $k_D$ and $k_S$:
\begin{equation}
\label{formula}
  p_{ij} = N\exp{\left(k_D D_{ij}\right)}
  \exp{\left(k_S S_{ij}\right)}p_I(i,j)p_W,
\label{trap}
\end{equation}
with the normalization $N$. Here $p_I$ represents the inertia
effect \cite{BKAJ} given by $p_I(i,j)=\exp{\left(k_I \right)}$ for
the direction of one's motion in the previous time step, and
$p_I(i,j)=1$ for other cells, where $k_I$ is the sensitivity
parameter. $p_W$, newly introduced in this paper, is the wall
potential which is explained below. In (\ref{trap}) we do not take
into account the obstacle cells (walls etc.) as well as occupied
cells.

\item Each pedestrian chooses randomly a target cell based on the transition
      probabilities $p_{ij}$ determined by (\ref{formula}).

\item Whenever two or more pedestrians attempt to move to the
      same target cell, the movement of {\em all}
involved particles is denied with probability $\mu \in [0,1]$,
i.e.\ all pedestrians remain at their site \cite{AKA}. This means
that with probability $1-\mu$ one of the individuals moves to the
desired cell. Which one is allowed to move is decided using a
probabilistic method \cite{BKAJ,AKA}.

\item The pedestrians who are allowed to move perform their motion
to the target cell chosen in step 3. $D$ at the origin cell
$(i,j)$ of each {\em moving} particle is increased by one:
$D_{ij}\to D_{ij}+1$, i.e.\ $D$ can take any non-negative integer
value.

\end{enumerate}
The above rules are applied to all pedestrians at the same time
(parallel update). Some important details are explained in the
following subsections.

\subsection{Effect of walls}
People tend to avoid walking close to walls and obstacles. This
can be taken into account by using ``wall potentials''. We
introduce a repulsive potential inversely proportional to the
distance from the walls. The effect of the static floor field is
then modified by a factor (see eq.~(\ref{trap})):
\begin{equation}
 p_W = \exp (k_W \min (D_{\rm max}, d)),
\end{equation}
where $d$ is the minimum distance from all the walls, and $k_W$ is
a sensitivity parameter. The range of the wall effect is
restricted up to the distance $D_{\rm max}$ from the walls.

\begin{figure}[tb]
\begin{center}
\includegraphics[width=0.22\textwidth]{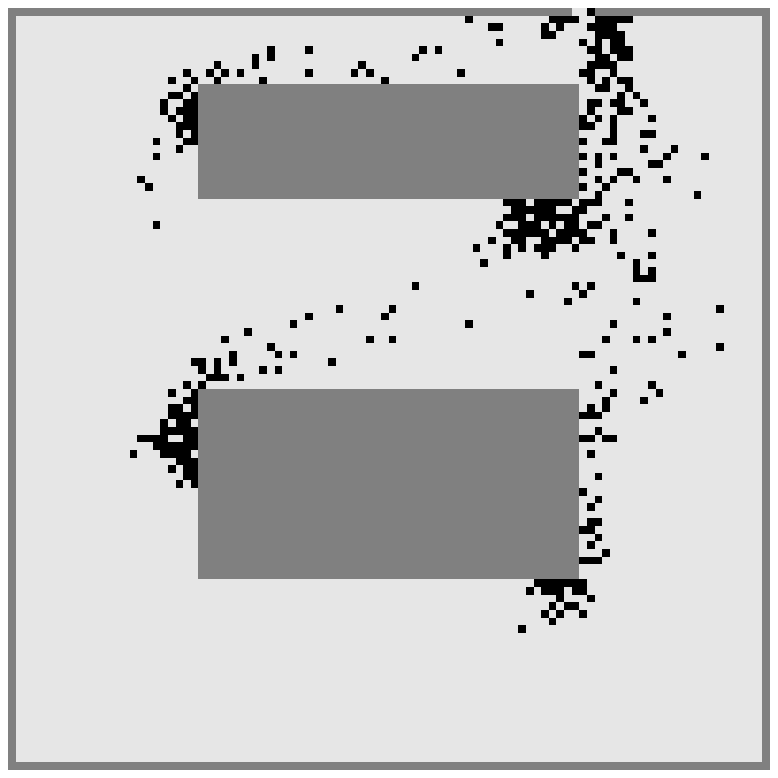}
\includegraphics[width=0.225\textwidth]{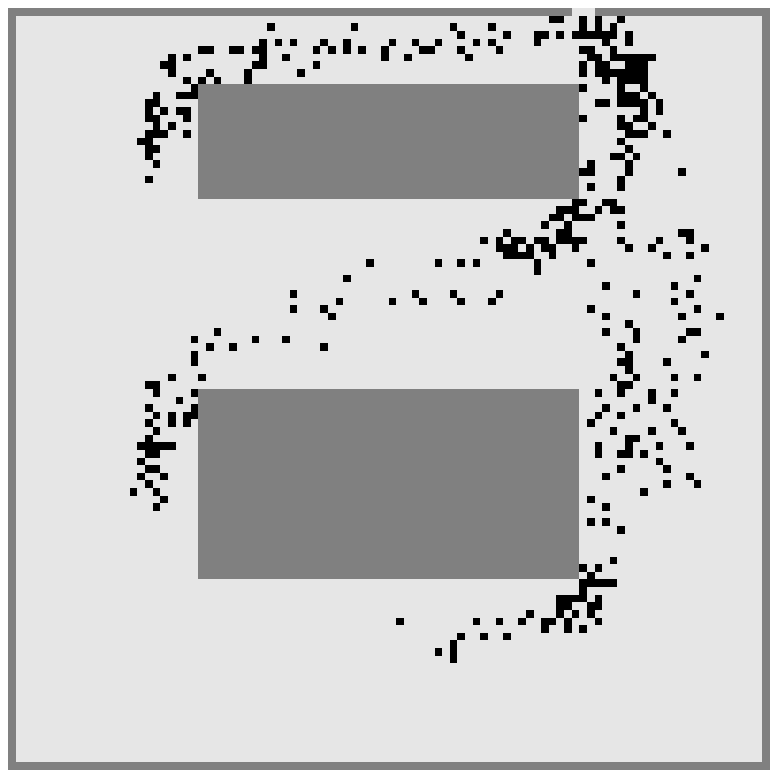}\\
(a)\hspace{3cm}(b)
\end{center}
\caption{
Snapshot of evacuation (a) without ($k_W=0$) and (b) with
($k_W=0.5$) wall potential.
We can clearly see the artifact of jamming at every corner
without the wall potential.
Parameters are $D_{\rm max}=10, k_S=2.0, k_D=1.0, k_I=0.2,
\mu=0.2$ and the initial density is $\rho=0.03$.
}
\label{figwall}
\end{figure}

Fig.~\ref{figwall} shows an example for an evacuation from a room
with obstacles using the wall potential. Without wall potentials
($k_W=0$), jamming areas near every corner can be observed,
because everybody tries to evacuate along the same path of minimum
length. For $k_W\neq 0$ these areas are clearly suppressed. Thus
the introduction of this additional potential improves the realism
of the model.
\subsection{Calculation of the static field in arbitrary geometries}
\begin{figure}[tb]
\begin{center}
\includegraphics[width=0.3\textwidth]{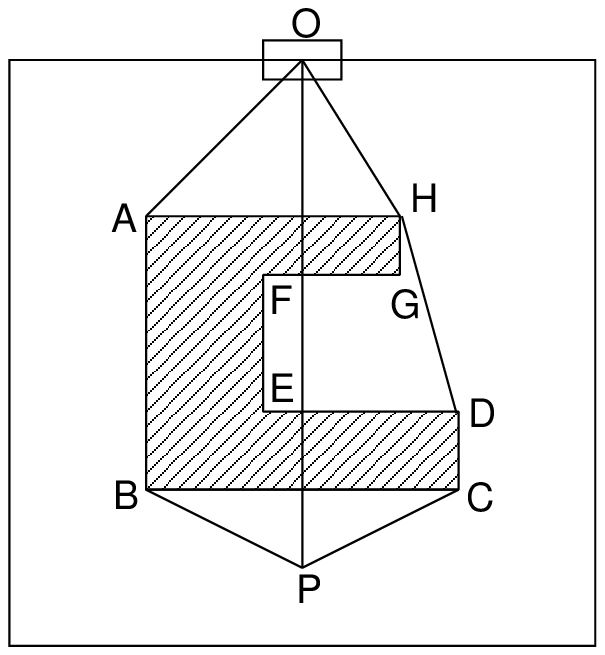}\\
(a)\\
\includegraphics[width=0.3\textwidth]{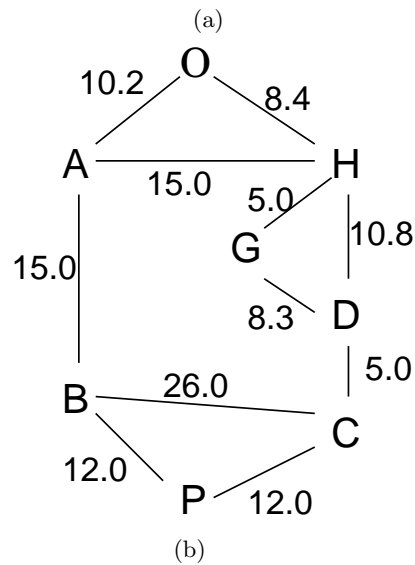}\\
\hspace{-0.5cm}(b)
\end{center}
\caption{
Example for the calculation of the static floor field using the Dijkstra
method. (a) A room with one obstacle. The door is at $O$ and the obstacle
is represented by lines $A-H$. (b) The visibility graph for this room.
Each node connected by a bond is ``visible'', i.e., there are no
 obstacles between them. The real number on each bond represents
the distance between them as an illustration.
}
\label{figdikstra}
\end{figure}
In the following we propose a combination of the visibility graph
and Dijkstra's algorithm to calculate the static floor field.
These methods enable us to determine the minimum Euclidian ($L^2$)
distance of any cell to a door with arbitrary obstacles between
them.

Let us explain the main idea of this method by using the
configuration given in Fig.~\ref{figdikstra}(a) where there is an
obstacle in the middle of the room. We will calculate the minimum
distance between a cell $P$ and the door $O$ by avoiding the
obstacle. If the line $PO$ does not cross the obstacle $A-H$, then
the length of the line, of course, gives the minimum. If, however,
as in the example given in Fig.~\ref{figdikstra}(a), the line $PO$
crosses the obstacle, one has to make a detour around it. Then we
obtain two candidates for the minimum distance, i.e., lines $PBAO$
and $PCDHO$. The shorter one finally gives the minimum distance
between $P$ and $O$.
If there are more than one obstacle in the room,
then we apply the same procedure to each of them repeatedly.
Here it is important
to note that all the lines pass only the obstacle's edges
with an acute angle. It is apparent that the obtuse edges like $E$
and $F$ can never be passed by the minimum lines.

To incorporate this idea into the computer program, we first need
the concept of the {\em visibility graph} in which only the nodes
that are visible to each other are bonded \cite{CG} (``visible''
means here that there are no obstacles between them). The set of
nodes consists of a cell point $P$, a door $O$ and all the acute
edges in the room. In the case of Fig.~\ref{figdikstra}(a), the
node set is $\{P,O,A,B,C,D,G,H \}$ and the bonds are connected
between $A-B$, $A-H$, and so on (Fig.~\ref{figdikstra}(b)). Each
bond has its own weight which corresponds to the Euclidian
distance between them.

Once we have the visibility graph, we can calculate the distance
between $P$ and $O$ by tracing and adding the weight of the bonds
between them. There are several possible paths between $P$ and
$O$, and the one with minimum total weight represents the shortest
route between them. The optimization task is easily performed by
using the Dijkstra method \cite{CG} which enables us to obtain the
minimum path on a weighted graph.

\begin{figure}[tb]
\begin{center}
\includegraphics[width=0.4\textwidth]{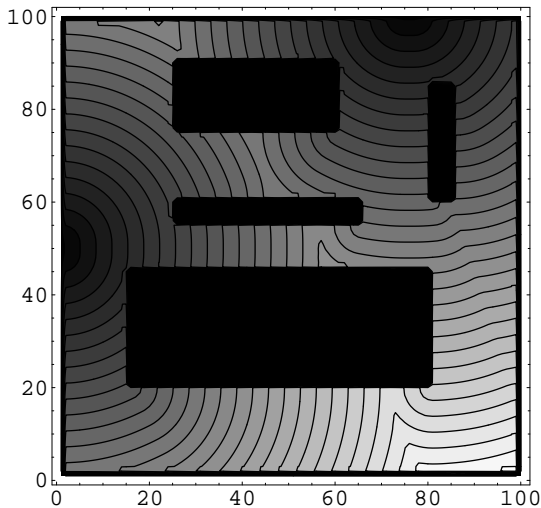}
\end{center}
\caption{A contour plot of the static floor field by using the
Dijkstra metric. There are four obstacles and two doors in this
room. The darkness of shading is inversely proportional to the
distance from the nearest door. } \label{figdik}
\end{figure}

Performing this procedure for each cell in the room, the method
allows us to
determine the static floor field for arbitrary geometries. We will
call this metric {\em Dijkstra metric} in the following. Results
for a complex static floor field obtained by this method are shown
in Fig.~\ref{figdik}. There are two doors in this room, thus we
calculate the minimum distance for each door from each cell in the
room and take the shorter one as value of the static floor field.

\begin{figure}[tb]
\begin{center}
\includegraphics[width=0.22\textwidth]{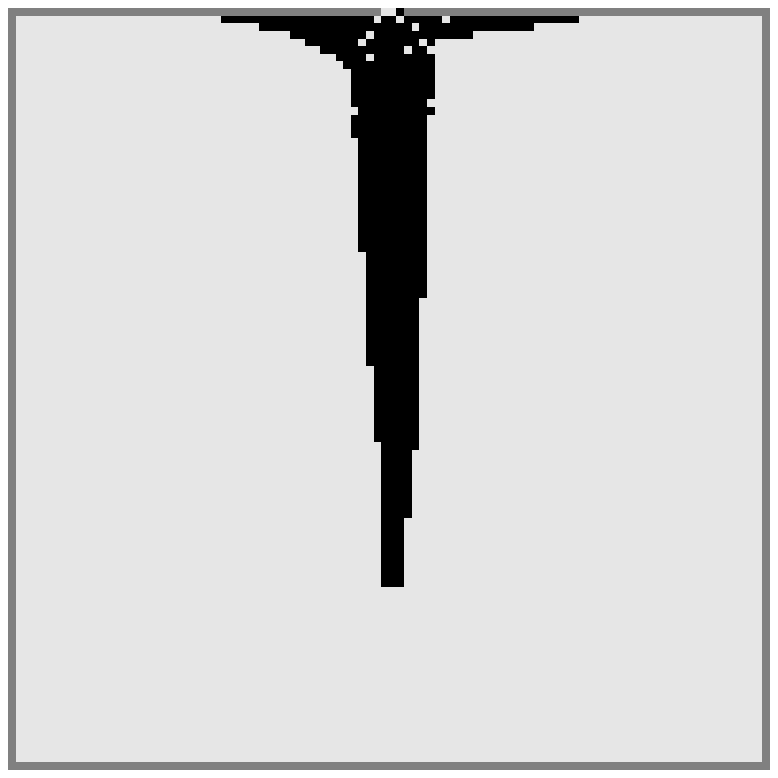}
\includegraphics[width=0.225\textwidth]{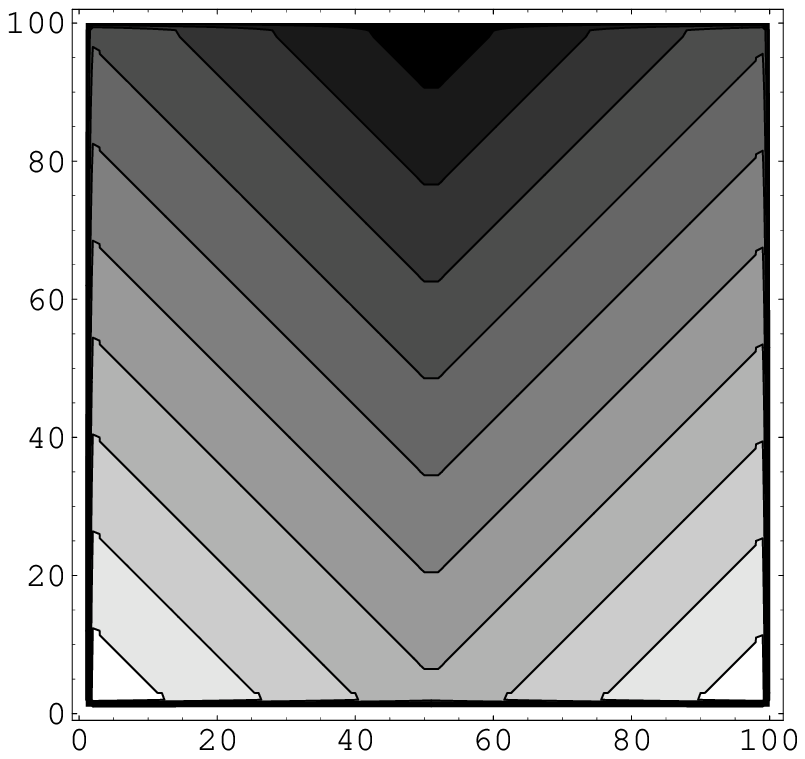}\\
(a)\hspace{3cm}(b)\\\vspace{0.5cm}
\includegraphics[width=0.22\textwidth]{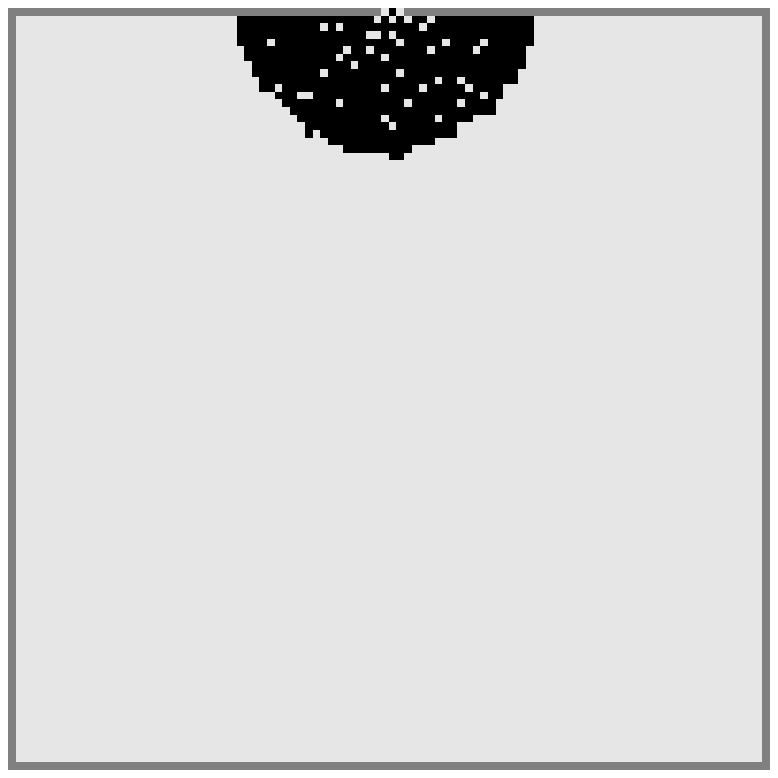}
\includegraphics[width=0.225\textwidth]{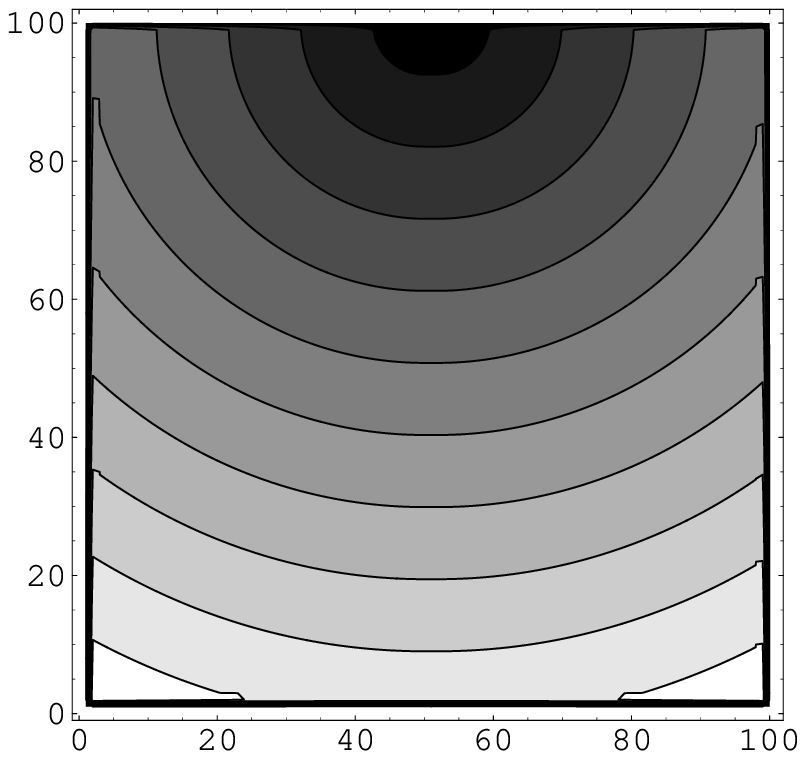}\\
(c)\hspace{3cm}(d)
\end{center}
\caption{Comparison of typical snapshots
in the case of the Manhattan (a) and Dijkstra (c) metric for the
static floor field.
We see an artifact of line formation in front of the door in (a).
This comes from the nonisotropic nature of
the Manhattan metric as seen in the contour plot
(b), while the Dijkstra metric gives
an isotropic static floor field from the door (d).
Parameters are $k_S=3.0, k_D=1.0, k_I=0.2,
\mu=0.2$ and the initial density is $\rho=0.04$.
}
\label{figcomp}
\end{figure}

Next, we like to point out the advantages of the Dijkstra metric
compared to the simpler Manhattan metric \cite{AA}.
Fig.~\ref{figcomp} shows a comparison of typical configurations
during an evacuation from a simple room with no obstacles by using
both the Dijkstra and the Manhattan metric. We see that for large
$k_S$ the pedestrians move preferably along a line in front of the
door if the Manhattan metric is used, whereas for the Dijkstra
metric the behavior is more realistic.

\subsection{Diffusion and decay of the dynamical floor field}
We can show that the order of diffusion and decay process in rule
1 (see Sec.~\ref{subrules}) is {\it exchangeable}, i.e., it makes no
difference no matter which of the two processes is applied first.
Both diffusion process and decay can be written in difference form
as
\begin{eqnarray}
 D^{t+1}_{ij}&=&D^t_{ij}-\alpha D^t_{ij}\nonumber\\
&+&\frac{\alpha}{4}(D^t_{i+1,j}+D^t_{i-1,j}
+D^t_{i,j+1}+D^t_{i,j-1}),\label{diff}\\
 D^{t+1}_{ij}&=&D^t_{ij}-\delta D^t_{ij},\label{decay}
\end{eqnarray}
respectively.
Thus the combination of the diffusion and decay above gives
\begin{eqnarray}
 D^{t+1}_{ij}&=&(1-\alpha)(1-\delta)D^t_{ij}
+\frac{\alpha(1-\delta)}{4}(D^t_{i+1,j}\nonumber\\
+&&+D^t_{i-1,j}+D^t_{i,j+1}+D^t_{i,j-1}) \label{decaydiff}
\end{eqnarray}
{\it regardless} of their order.

In \cite{BKAJ} also a continuous dynamic floor field has been
investigated. It was observed that this has no qualitative
influence on the behavior of the model.
This comes from the fact that
the continuum limit of
(\ref{decaydiff}) is the same as the continuous version in \cite{BKAJ}.
The limit of (\ref{decaydiff}) is given by
\begin{equation}
 \frac{\partial D}{\partial t}=-d_1 D
+\frac{\alpha d_2}{4}\left(\frac{\partial^2 D}{\partial x^2}+
\frac{\partial^2 D}{\partial y^2}\right),\label{Cdecaydiff}
\end{equation}
where $\delta / \Delta t =d_1$ and $(\Delta x)^2 / \Delta
t=(\Delta y)^2 / \Delta t=d_2$ (time and space intervals are
written as $\Delta t$ and $\Delta x$, $\Delta y$), respectively,
and $d_1, d_2$ and $\alpha$ are kept constant in the limit $\Delta
x, \Delta y, \Delta t, \delta \to 0$.
Therefore the
dynamics of $D$ given by (\ref{Cdecaydiff}) coincides with the
previous one if we choose appropriate coefficients.
\subsection{Model parameters and their physical relevance}
There are several parameters in our model, and the most important
ones are listed below with their physical meaning which is helpful
in understanding the collective behavior in the simulations.
\begin{enumerate}
\item $k_S\in [0,\infty[ \cdots $ The coupling to the static field
characterizes the knowledge of the shortest path to the doors, or
the tendency to minimize the costs due to deviation from a planned
route \cite{Ho}. This considerably controls one's velocity and
evacuation times.
 \item $k_D\in [0,\infty[ \cdots $ The coupling to the dynamic
field characterizes the tendency to follow other people ({\em
herding behavior}). The ratio $k_D/k_S$ may be interpreted as the
degree of panic. It is known that people try to follow others
particulary in panic situations \cite{HFV} (see
Sec.~\ref{humandata}). This tendency lasts at least until they can
escape without any hindrance. If hindrance by other people takes
place often and a tendency to clogging emerges, people try to
avoid such directions. This is taken into account in our dynamic
floor field which is proportional to the velocity density, such
that people will follow only {\em moving} persons.

 \item $k_I\in [0,\infty[ \cdots $ This parameter determines the
 strength of inertia which suppresses quick changes of the direction of motion.
It also reflects the tendency to minimize the costs due to
deviation from one's desired route and acceleration \cite{Ho}.

 \item $\mu\in[0,1]\cdots $ The friction parameter controls
 the resolution of conflicts in clogging situations. Both
cooperative and competitive behavior at a bottleneck are well
described by adjusting  $\mu$ \cite{AHKAM}.

 \item $\alpha, \delta\in[0,1]\cdots $ These constants control
diffusion and decay of the dynamic floor field. It reflects the
randomness of people's movement and the visible range of a person,
respectively. If the room is full of smoke, then $\delta$ takes
large value due to the reduced visibility. Through diffusion and
decay the trace is broadened, dilute and vanishes after some time.

 \item $k_W, D_{\rm max}\cdots $ These parameters specify the wall
 potential. Pedestrians tend to avoid walking close to walls and obstacles.
$D_{\rm max}$ is the maximum distance at which people feel the
walls. It reflects one's range of sight or so-called personal
space \cite{space}. $k_W$ is the sensitivity to the walls, and the
ratio $k_W/k_S$ reflects to which degree deviations from the
shortest route (which is determined by the static floor field) are
accepted to avoid the walls.
\end{enumerate}
\section{Simulations}\label{simulation}
We focus on measuring the total evacuation time by changing the
parameters $k_S, k_D, k_I, \mu$ and the configuration of the room,
such as width, position and number of doors and obstacles. In all
simulations we put $D_{\rm max}=10$, $\alpha=0.2$ and
$\delta=0.2$, and von Neumann neighborhoods are used in
eq.~(\ref{trap}) for simplicity. The size of the room is set to
100 $\times$ 100 cells.

In the previous papers \cite{BKAJ,AA,AKA}, the influence of the
two floor fields on the total evacuation time has been studied in
detail. Here, the effects of inertia and wall potentials are
investigated for concave rooms with some obstacles by using the
Dijkstra metric.
\subsection{Inertia effect}

\begin{figure}[tb]
\begin{center}
\includegraphics[width=0.45\textwidth]{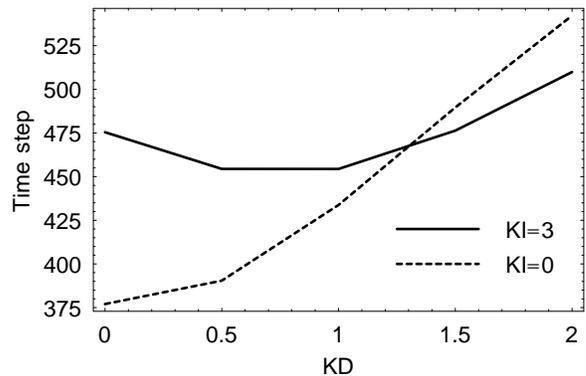}
\end{center}
\caption{
Total evacuation time versus coupling $k_D$ to the dynamic
floor field in the dependence of $k_I$.
The room is a simple square without obstacles and 50 simulations are
averaged for each data point.
Parameters are $\rho=0.03, k_S=2, k_W=0.3$ and $\mu=0$.}
\label{figinertia}
\end{figure}

Pedestrians try to keep their preferred velocity and direction as
long as possible. This is taken into account by adjusting the
parameter $k_I$. In Fig.~\ref{figinertia}, total evacuation times
from a room without any obstacles are shown as function of $k_D$
in the cases $k_I=0$ and $k_I=3$. We see that it is monotonously
increasing in the case $k_I=0$, because any perturbation from
other people becomes large if $k_D$ increases, which causes the
deviation from the minimum route. Introduction of inertia effects,
however, changes this property qualitatively as seen in
Fig.~\ref{figinertia}.
The {\it minimum} time appears around $k_D=1$ in the case $k_I=3$.
This is well explained by taking into account
the physical meanings of $k_I$ and $k_D$.
If $k_I$ becomes large,
people become less flexible and all of them try to keep their own
minimum route to the exit according to the static floor field
regardless of congestion.
By increasing $k_D$, one begins to feel the disturbance from other people
through the dynamic floor field. This perturbation
makes one flexible and hence
contributes to avoid congestion.
Large $k_D$ again works as strong
perturbation as in the case of $k_I=0$, which diverts people
from the shortest route largely.
Thus we have the minimum
time at a certain magnitude of $k_D$, which will depend on the
value of $k_S$ and $k_I$.

\subsection{Contraction at a wide exit}
\begin{figure}[tb]
\begin{center}
\includegraphics[width=0.22\textwidth]{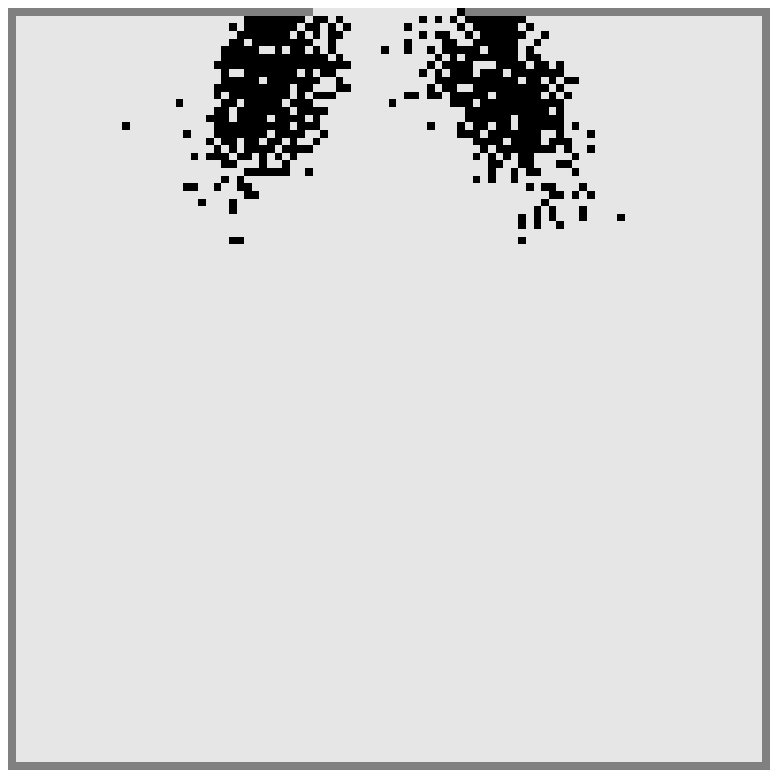}
\includegraphics[width=0.22\textwidth]{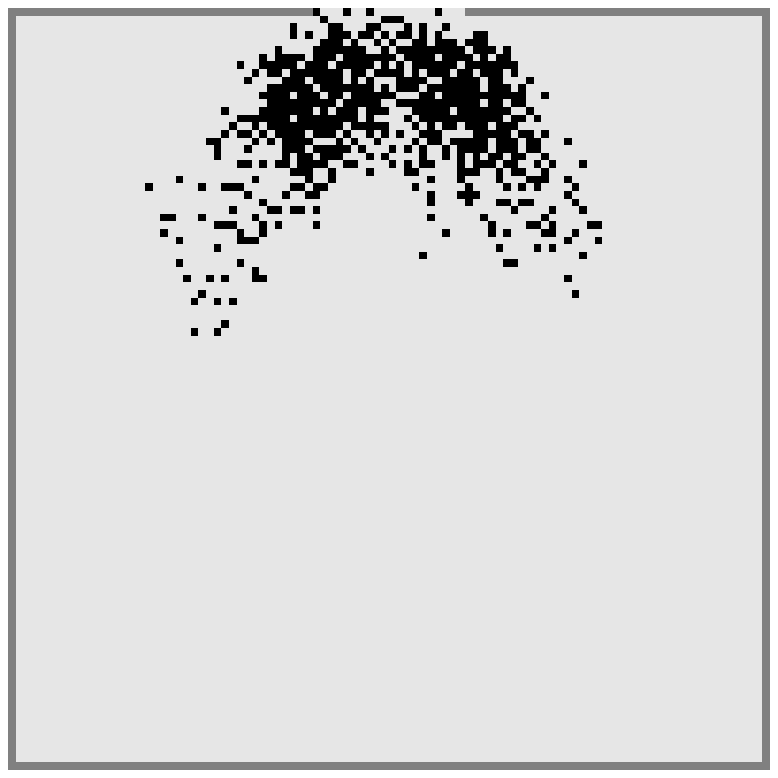}\\
(a)\hspace{3cm}(b)
\end{center}
\caption{
Contraction of flow through a wide exit. The width of the exit
is set as 20 cells. In (a) we set the ratio of contraction as 1, while
in (b) it is 0.3.
In (a) we see the artifact of the
crowd at both ends of the exit even if they can
easily evacuate through the center of the exit.
Parameters are $\rho=0.05, k_I=1.0, k_S=2.0, k_D=1.0,
k_W=0.3$ and $\mu=0.2$.
}
\label{figcontra}
\end{figure}

If the width of an exit becomes large, a more careful treatment is
needed in the calculation of the static floor field. People tend
to rush to the center of the exit to avoid the walls. Thus one
should introduce an effective width of the exit by neglecting
certain cells from its each end. We call this effect {\it
contraction} in this paper, due to its similarity with the
contraction effect in hydrodynamics where fluid runs through a
orifice with a smaller diameter than that of the orifice
immediately after the fluid goes out of it \cite{fluid}. The
shortest distance from a cell in the room to one of the exit cells
is calculated by using the Dijkstra metric, but only those exit
cells near the center of the door are taken into account owing to
the contraction. Then we take the minimum of those shortest
distances and use it as the value of the static floor field at the
cell. Here we define the ratio of contraction of an exit as
$c=W'/W$, where $W$ is the true width of the exit and $W'$ is the
effective width. If $c=1$, i.e., there is no contraction, and we
see the artifact of two crowds near the edges of the exit
(Fig.~\ref{figcontra}(a)). Introducing the contraction makes the
evacuation behavior more realistic (Fig.~\ref{figcontra}(b)).
\subsection{Effect of obstacles}\label{posob}

\begin{figure}[tb]
\begin{center}
\includegraphics[width=0.22\textwidth]{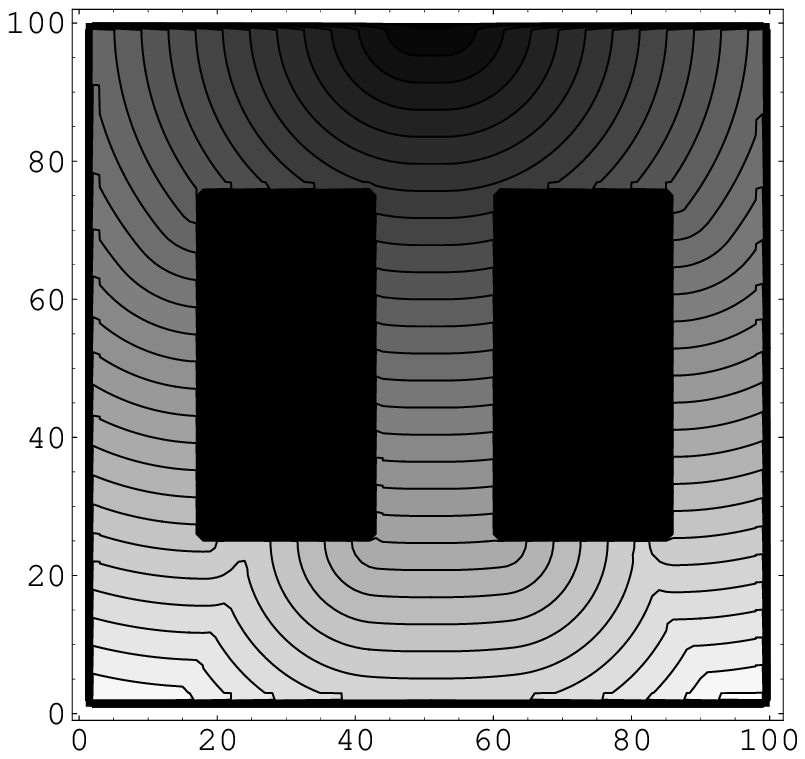}
\includegraphics[width=0.22\textwidth]{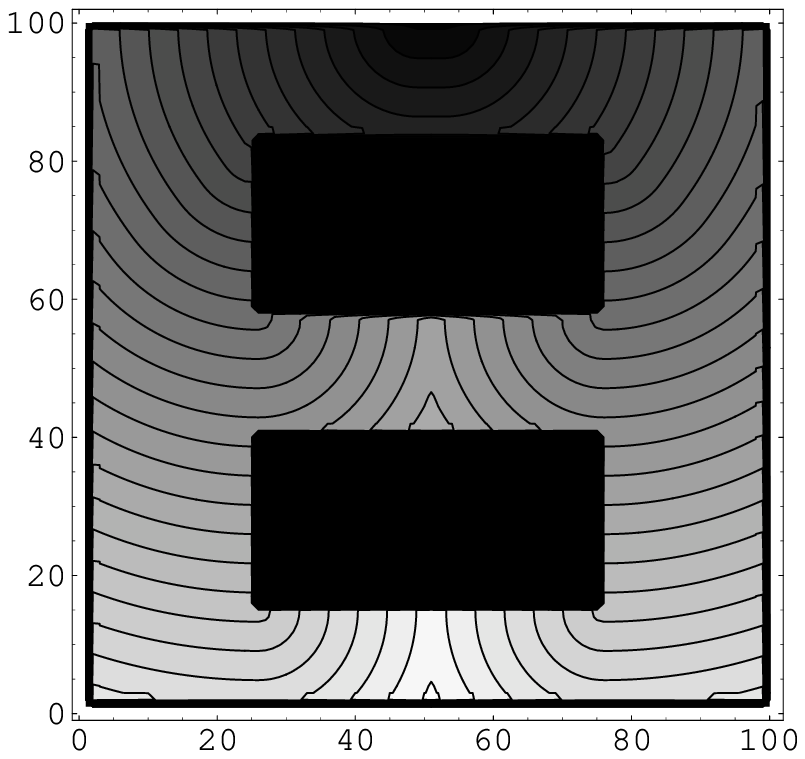}\\
(a)\hspace{3cm}(b)
\end{center}
\caption{ The static floor field of rooms with same obstacles
arranged in a different way. Although the total area of obstacles
is the same, the maximum length to the exit in the room (a) is
115.5, while 124.6 in (b). Simulations are done with parameters
$k_S=2.0, k_D=1.0, k_I=3.0, k_W=0.3, \mu=0.2$ and the initial
density of the room is $\rho=0.03$. } \label{figobst}
\end{figure}

Let us investigate the effect of the position of obstacles to the
total evacuation time. In Fig.~\ref{figobst}, we set up two rooms
such that the total area of obstacles in both rooms is the same.
However, it is important to notice that the maximum length to the
exit is different for these rooms. The maximum length in room (b)
is 124.6, which is longer than that in (a) (115.5). This
difference affects the static floor field and hence the dynamics
of people. The average total evacuation time in case (a) is given
by 357.48 time steps, while in (b) it is 379.22 time steps. This
implies that, even though the area of obstacles is the same, their
positions in the room will affect the evacuation dynamics
considerably through the static floor field. It is worth
mentioning that this simulation is different from the column
problem we have studied previously \cite{AKA}. The existence of a
small column in front of an exit does not change the static floor
field so much, but works as a simple obstacle which divides the
flow of evacuating people.

\subsection{Influence of exit width and number of doors}

\begin{figure}[tb]
\begin{center}
\includegraphics[width=0.22\textwidth]{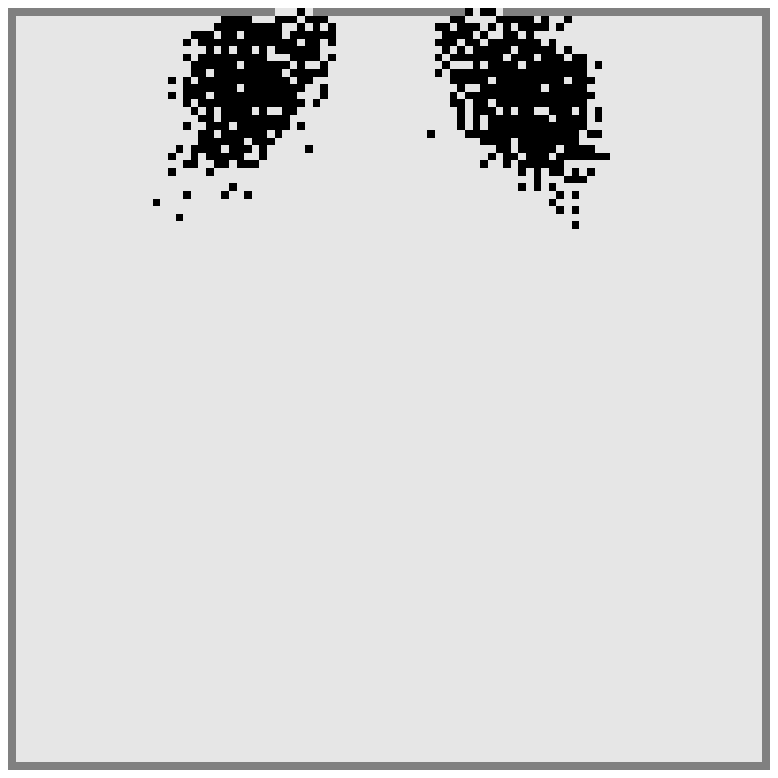}
\includegraphics[width=0.22\textwidth]{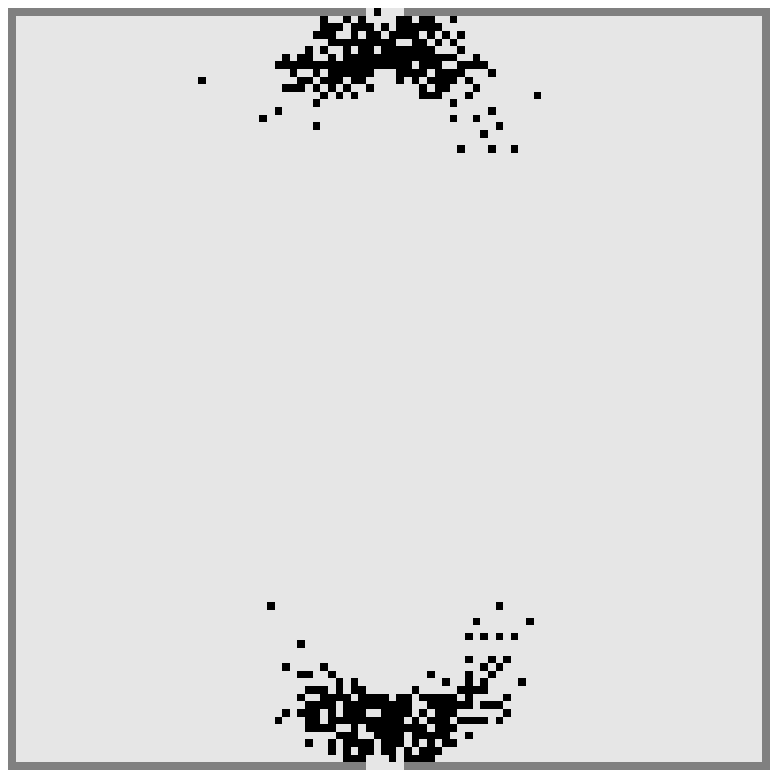}\\
(a)\hspace{3cm}(b)
\end{center}
\caption{ A snapshot of an evacuation from a room with two exits.
The total width of the exits is the same as for the one exit case.
Simulations are done with parameters $k_S=2.0, k_D=1.0, k_I=1.0,
k_W=0.3, \mu=0.2$ and $\rho=0.03$. } \label{figexit}
\end{figure}

Finally we study the effect of the width of an exit as well as the
total number of doors in a room. We compare a room with an exit of
size 10 to one with two exits of size 5 (Fig.~\ref{figexit}).
Although the total width of the exits is the same, the evacuation
dynamics is different. The total evacuation time is 275 time steps
in average for the case of one exit, but 245 for two exits
(Fig.\ref{figexit}(a)). If the two doors are set at opposite
walls, the evacuation time is further improved to 220 time steps
(Fig.\ref{figexit}(b)). This is similar to the effect studied in
Sec.~\ref{posob}, because the minimum length in the case of
Fig.~\ref{figexit} (b) becomes 68.2 while it is 104.1 in (a).

\section{Concluding discussion}\label{conc}

In this paper we have discussed the main properties of the floor
field cellular automaton model for pedestrian dynamics. A method
for the calculation of the static floor field and the introduction
of wall potentials has been proposed. Both extensions improve the
realism of evacuation simulations. Also it is important to take
into account the contraction in the case of wide exits. The
existence of a minimal evacuation time in the case of finite
inertia $k_I\ne 0$ is found, which shows the importance of one's
flexibility to respond to the other people's behavior in the case
of an evacuation.
Finally it is shown that the position of obstacles in a room will
affect the evacuation dynamics through the static floor field.
Thus in architectural
planning it is important to consider a suitable position of
obstacles in a room as well as the exits carefully.

\section*{Acknowledgement}
This work was supported in part by the Ryukoku fellowship
 2002 (K. Nishinari).

\end{document}